\documentstyle[12pt,epsf]{article}
\def\eps {\epsilon}
\begin{document}
\thispagestyle{empty}
\normalsize
%
\begin{center}
\vspace{3cm}
\large\bf{POLARIZATION IN SEMILEPTONIC 
$B\rightarrow X\tau$ 
DECAYS
\footnote{Work supported in part by KBN grant 2P03B08414
and by BMBF grant POL-239-96.}} \\
\vspace{1cm}
{\large\bf Marek Je\.zabek$^{a,b}$ and Piotr Urban$^{a}$} \\
\vspace{0.5cm}
{\normalsize\it  
 $^a$\ Institute of Physics, University of Silesia,
     ul. Uniwersytecka 4,\\
  PL-40007 Katowice, Poland\\
 $^b$\ Institute of Nuclear Physics, Kawiory 26a, PL-30055 Cracow, Poland} 
\end{center}
\vspace{3cm}
\begin{abstract}
The paper gives the polarization of the tau lepton in the semileptonic B 
decays with respect to the direction of the virtual W boson. The result 
is given including the nonperturbative HQET corrections. The perturbative 
QCD corrections are probably negligible as suggested by the existing 
results for the longitudinal polarization of the charged lepton (Je\.zabek 
and Urban, 1998). 
\end{abstract} \vspace{4.cm}
\vspace{1.cm}
\pagebreak
\setcounter{page}{0}
\section{Introduction}
The interest in semileptonic B decays is currently increasing as the B 
factory in KEK is scheduled to begin to collect data later this year. 
This domain of physics is likely to upgrade our knowledge on the 
Standard Model parameters as well as to provide tests on its validity. 
The semileptonic B decays can contribute to the former as their 
theoretical description is now far more successful than that of the 
hadronic processes\cite{PW}--\cite{Georgi}.
\par  
The polarization of the charged lepton does not depend on the 
Cabibbo-Kobayashi-Maskawa matrix element and so it can be instrumental 
in finding the quark masses.  The longitudinal polarization of the tauon 
including first order perturbative QCD corrections
has been found analytically \cite{JU} 
by taking the analytical decay width for the unpolarized case \cite{JM} 
and calculating the width for a negative polarization. 
Then the result can be integrated to give polarized
tau energy distribution. The method used in that calculation can easily be modified to give other polarizations. This fact matters insomuch that 
experimentally it is the polarization along the intermediating 
$W$ boson direction that is easier to measure \cite{Rozanska}, see also
\cite{KiSo}. 
The reason is that the direction of $\tau$ lepton can be determined
at B factories with rather poor accuracy. On the other hand the direction
of $W$ is opposite to the direction of hadrons in semileptonic B decays.
The latter can be well measured at least for the exclusive 
$B\to D\tau\bar\nu_\tau$ and $B\to D^*\tau\bar\nu_\tau$ channels
which probably contribute the dominant contribution to the inclusive
decay rate.

The present calculation includes tree level and HQET corrections only.
We are unable to calculate perturbative QCD corrections because the
analytic structure of expressions is far more complicated than in the
case of the longitudinal polarization\cite{JU}. However, 
indications exist that the effects of perturbative QCD corrections 
on $\tau$ polarization are negligible. 
In particular, the above-mentioned 
longitudinal polarization does not change visibly after the first-order 
perturbative corrections have been included 
either in the rest frame of the $W$ boson \cite{CJK} or that of the 
decaying quark \cite{JU}.

The paper is broken up into four sections. In Sec.2, kinematical 
variables are introduced. Secs.3 and 4 are to explain the method used in
the calculation and then the results are shown is Sec.5. In Appendix A
some details of HQET calculations are explained including the discussion
of singularity problems. Such problems were also encountered in 
\cite{GKS} where a method was proposed to eliminate them.

\section{Kinematical variables}
In this section we define the kinematical variables used throughout the article
as well as their boundaries. The calculation is performed in the 
rest frame of the decaying $ B$ meson, which coincides with that of the $b$ quark at the tree level in the parton model.
The four-momenta of the particles are denoted as following: $Q$ for the 
$b$ quark, 
$q$ for the $c$ quark, $W$ for the virtual $W$ boson, $\tau$ for 
the charged lepton, and $\nu$ for the corresponding 
antineutrino. All the particles are assumed to be on-shell so that their
squared four-momenta equal their masses:
\begin{equation}
Q^{2}=m^{2}_{b}\ ,\quad \ q^{2}=m^{2}_c\ ,\quad 
\tau^{2}=m^{2}_{\tau}\ ,\quad \nu^{2}=0 \ .
\end{equation}
The employed variables are scaled to the
units of the decaying quark mass $m_b$:
\begin{equation}\label{variables}
\rho={{m_c^2} \over {m_b^2}}\ ,\quad \eta={{m_{\tau}^2} \over {m_b^2}}\ ,
\quad 
y={{2E_{\tau}} \over {m_{b}}}\ ,\quad t={{W^2} \over {m_{b}^{2}}}\ ,\ 
x={{2E_{\nu}} \over {m_b}}\ .
\end{equation}
Henceforth we scale all quantities so that $m_b^2=Q^2=1$. The charged lepton
 is described by the light-cone variables:
\begin{equation}
\tau_{\pm}={1 \over 2}\left(\, y \pm \sqrt{y^2-4\eta}\,\right).
\end{equation}
The $W$ boson is characterized likewise:
\begin{eqnarray}
w_0={1\over 2}(1+t-\rho)\ ,\\
w_3=\sqrt{w_0^2 - t}\ ,\\
w_{\pm}=w_0 \pm w_3\ .
\end{eqnarray}
The phase space is defined by the ranges of the kinematical variables:
\begin {eqnarray}
2\sqrt{\eta}&\le y \le & 1+\eta-\rho=y_m\quad ,\\
t_{min}={\tau}_- \left(1-{\rho \over {1-{\tau}_-}}\right) &\le t\le &
{\tau}_+\left( 1-{\rho \over 1-{\tau}_+}\right)=t_{max}\quad .
\end{eqnarray}
The limits above are obtained within the parton model approximation.
They change if we allow for Fermi motion, which we must in order to be
able to discuss the HQET corrections to the decay 
widths\cite{GKS}--\cite{FLNN22}. 
Also, contrary to
the parton model case, the energy of neutrino can vary within limits 
which depend in a non-trivial manner on the values of the 
variables $y,t$. The details of the subject 
were discussed in \cite{BKPS24},
so we will only state here that the integrations involving delta functions 
and their derivatives have the effect of confining the range of the variables
$y,t$ to that of the parton model.

\section{Polarization evaluation}
The polarization is found by evaluating the unpolarized decay width and 
any of the two corresponding to a definite polarization, 
according to the definition,
\begin{equation}
P={{\Gamma ^+-\Gamma ^-}\over{\Gamma ^++\Gamma^-}}
=1-2{{\Gamma ^-}\over{\Gamma}}\quad ,
\end{equation}
where $\Gamma=\Gamma ^++\Gamma ^-$. 
The calculation of the polarized width is structured after the manner of 
that which has yielded the longitudinal polarization \cite{JU}. 
Thus in the rest frame of the decaying quark, one can decompose:
\begin{equation}\label{stolQ}
s={\cal A}Q +{\cal B}W \quad .
\end{equation}
The coefficients ${\cal A},{\cal B}$ can be evaluated using the conditions 
defining the polarization four-vector $s$, which reduce to the following 
when the parton model value of the neutrino energy is assumed:
\begin{eqnarray}
{\cal A}_0^{\pm}&=&\mp {{t+\eta}\over\sqrt{t(y-y_-)(y_+ - y)}}\quad ,\\
{\cal B}_0^{\pm}&=&\pm {y\over\sqrt{t(y-y_-)(y_+ - y)}}\quad , 
\end{eqnarray}
where the superscripts at ${\cal {A,B}}$ denote the polarization of the 
lepton,while
\begin{equation}
y_{\pm}=(1+\eta /t)w_{\pm} \quad .
\end{equation}
This observation is made relevant by the fact that the decay width for a 
definite polarization of the charged lepton is gotten from the analogous 
expression for the unpolarized case,
\begin{equation}
d\Gamma_0=G_F^2M_b^5|V_{CKM}|^2{\cal M}_{0,3}^{un}
d{\cal R}_3(Q;q,\tau,\nu)/\pi^5 
\end{equation}
where
\begin{equation}\label{tauliniowy}
{\cal M}_{0,3}^{un}(\tau)= q\cdot\tau Q\cdot\nu\quad , 
\end{equation}
by formally replacing the lepton four-momentum by the following 
four-vector $K$:
\begin{equation}
K=\tau -m_{\tau}s \quad.
\end{equation}
Then we obtain, 
\begin{equation}\label{Kliniowy}
{\cal M}_{0,3}^{pol}={1\over 2}{\cal M}_{0,3}^{un}(K=\tau -ms)=
{1\over 2}(q\cdot K) (Q\cdot\nu) \quad . 
\end{equation}
Although the expressions above are written for the Born approximation, 
corrections received by the hadronic tensor obviously do not alter this
scheme, so that we can apply it to the HQET calculations, too. However, 
the coefficients in the decomposition (\ref{stolQ}) need to be re-evaluated, 
taking into account the Fermi motion and working in the rest frame of 
the decaying meson. For a derivation of these cf. Appendix A.
Applying now the representation (\ref{stolQ}) of the polarization four-vector
$s$ we readily obtain the following useful formula for 
the matrix element with the lepton polarized:
\begin{equation}\label{abexp}
{\cal M}^{\pm}={1\over 2}{\cal M}^{un}(\tau)
\mp {{\sqrt{\eta}}\over{2\sqrt{y(t+\eta)(x+y)-y^2t-(t+\eta)^2}}}
    \left[y{\cal M}^{un}(W)-(t+\eta){\cal M}^{un}(Q)\right]\ .
\end{equation}
The above expression is valid for the HQET corrections, too.
The first term on the right hand side of (\ref{abexp}) can be calculated 
immediately once we know the result for the unpolarized case. Then the 
other terms require the formal replacement of the four-momenta 
$\tau \rightarrow W$ and $\tau \rightarrow Q$ in the argument.
\section{Evaluation of HQET corrections}
Using the operator expansion technique, one can obtain corrections to 
the decay widths of heavy hadrons which effectively lead to new terms in
the
hadronic tensor appearing in the triple differential decay width,
\begin{equation}\label{d3}
{{d\Gamma}\over{dx dt dy}}={{|V_{cb}|^2 G_F^2}\over{2\pi ^3}}
{\cal L}_{\mu\nu}{\cal W}_{\mu\nu} \quad .
\end{equation}
The hadronic tensor ${\cal W}$, related to an inclusive decay of a beautiful
hadron $H_b$,
\begin{equation}
{\cal W}_{\mu\nu}=(2\pi)^3\sum_{X}\delta ^4(p_{H_b}-q-p_X)
<H_b(v,s)|J_{\mu}^{c\dagger}|X><X|J_{\nu}^c|H_b(v,s)>
\end{equation}
can be expanded in the form
\begin{equation}\label{Wn}
{\cal W}_{\mu\nu}=-g_{\mu\nu}W_1+v_{\mu}v_{\nu}W_2-i\eps_{\mu\nu\alpha\beta}
v^{\alpha}q^{\beta}W_3+q_{\mu}q_{\nu}W_4
+(q_{\mu}v_{\nu}+v_{\mu}q_{\nu})W_5\ .
\end{equation}
The form factors $W_n$ can be determined by using the relation between the 
tensor ${\cal W}$ and the matrix element of the transition operator 
\begin{equation}
T_{\mu\nu}=-i\int d^4xe^{-iqx}T[J_{\mu}^{c\dagger}(x)J_{\nu}^c(0)]\ ,
\end{equation}
which is
\begin{equation}
{\cal W}_{\mu\nu}=-{1\over\pi} Im <H_b|T_{\mu\nu}|H_b>\ .
\end{equation}
The coefficients $W_n$ of (\ref{Wn}) have all been found elsewhere, see
eg. \cite{BKPS24} for a complete list.
Then the distribution (\ref{d3}) can be schematically cast in the following
form:
\begin{equation}\label{d3a}
{{d\Gamma}\over{dx dt dy}}=
f_1\delta(x-x_0)+f_2\delta '(x-x_0)+f_3\delta ''(x-x_0)\ ,
\end{equation}
where
\begin{equation}
x_0=1+t-\rho-y
\end{equation}
is the value of the neutrino energy in the parton model kinematics.
The triple differential distribution must be integrated 
over the neutrino energy to give
meaningful results. The final lepton energy distribution obtained on two 
subsequent integrations may be trusted except for the endpoint region where 
the operator product expansion fails.
In the present paper we give the double differential distribution so that
the lepton energy distribution has to be obtained numerically. The calculation
does not show any features unfamiliar from the cases of the other known 
polarizations. 
\section{Results}
\subsection{Double differential distribution} 
The polarized distribution can be written in the form,
\begin{equation}\label{HQET2}
{1\over\Gamma_0}\;
{{d\Gamma ^{\pm}}\over{\,dy\,dt}}={1\over2}F^{un}\pm
\left(\widetilde{F}
+\widetilde{F}_+ - \widetilde{F}_-\right)\ ,
\end{equation}
  where
\begin{equation}\label{G0}
{\Gamma}_0={{G_F^2m_b^5}\over{192{\pi}^3}}|V_{CKM}|^2 \ .
\end{equation}
The first term on the right hand side of Eq.(\ref{HQET2}) stands 
for the unpolarized
distribution, given e.g. in \cite{BKPS24}, Eq.$(30)$. 
Here we will only give  the new other term:\footnote{A 
FORTRAN 
code for this formula 
is available from piotr@charm.phys.us.edu.pl } 
\begin{eqnarray}\label{LW2}
\widetilde{F}&=& \sqrt{\eta}\;{\cal W}\,
\left[ 6 f_1 + {K_b}{\cal W} 
\left( f_2 + f_3 {\cal W}^2 +
{\textstyle{3\over 2}} f_4 {\cal W}^4 \right)
+ {G_b} \left( f_5 + f_6 {\cal W}^2 \right)  \right]\ ,
\end{eqnarray}
where
\begin{eqnarray}
f_1 &=& - yt (   1 + \rho - \eta ) + 2 yt^2 
- y (   1 + \rho\eta - 2\rho + \rho^2 + \eta ) - y^2t 
+ y^2  ( 1 - \rho )
\nonumber  \\&&
 +\, t  ( 1 + 2\rho\eta - \rho^2 )
+ t^2  ( 2\rho - \eta ) - t^3 - \rho^2\eta + \eta \ ,
  \\
f_2 &=& - 8yt  (  1 - \rho + \eta ) -16yt^2 
+8 y  ( 1 + \rho\eta - 2\rho + \rho^2 - \eta ) + 6y^2t 
\nonumber  \\&&
-\, 6y^2( 1- \rho ) + 2t ( 1 - 6\rho\eta 
- 4\rho + 3\rho^2 + 12\eta )
+ 2t^2  ( 8 - 6\rho + 3\eta ) 
\nonumber  \\&&   
+\, 6t^3
- 8\rho\eta + 6\rho^2\eta + 2\eta + 8\eta^2 \ ,
  \\
f_3 &=& yt  (  - 10\rho\eta + 6\rho\eta^2 
               + 14\rho^2\eta + 9\rho^2\eta^2 
               - 6\rho^3\eta + 2\eta 
                + 9\eta^2 - 4\eta^3 )
\nonumber  \\&&
      +\, yt^2  ( 1 - 9\rho\eta^2 - 5\rho 
                + 18\rho^2\eta + 7\rho^2 
                - 3\rho^3 + 6\eta - 13\eta^2 )
\nonumber  \\&&
       +\, yt^3  ( 1 - 18\rho\eta - 2\rho + 
                  9\rho^2 - 14\eta + 3\eta^2 )
       - yt^4  (  5 + 9\rho - 6\eta )
\nonumber  \\&&
    +\, 3yt^5    + y  (  - 5\rho\eta^2 + 4\rho\eta^3 
              + 7\rho^2\eta^2 - 3\rho^3\eta^2 
                  + \eta^2 + 4\eta^3 )
\nonumber  \\&&
       +\, y^2t  ( 1 + 14\rho\eta + 2\rho\eta^2 
                  - 3\rho - \rho^2\eta + 3\rho^2 
                  - \rho^3 - 13\eta + 8\eta^2 )
\nonumber  \\&&
       +\, y^2t^2  (  - 6 + 5\rho\eta 
                    + 6\rho + 15\eta - \eta^2 )
       + y^2t^3  ( 7 + 3\rho - 3\eta )
        - 2y^2t^4
\nonumber  \\&&
        +\, y^2 \eta  (  - 3\rho + 8\rho\eta
                + 3\rho^2 -  \rho^2\eta
                - \rho^3  - 7\eta + 1 )
        + y^3t  ( 1 - \rho )^2 
        - y^3t^2  \eta 
\nonumber   \\&&
        -\,  y^3t^3 
        + y^3 \eta  ( 1 - \rho )^2 \ ,
\\
f_4 &=& y^2t  (  - 6\rho\eta^2 - 2\rho^2\eta^3 
          + 6\rho^3\eta^2 + 4\rho^3\eta^3 
          - 3\rho^4\eta^2 + 3\eta^2 - 2\eta^3 )
\nonumber  \\&&
    +\, y^2t^2  (  - 6\rho\eta - 2\rho\eta^3 - 6\rho^2\eta^2 
             - 6\rho^2\eta^3 + 6\rho^3\eta + 12\rho^3\eta^2 
             - 3\rho^4\eta + 3\eta - 6\eta^2  ) 
\nonumber  \\&&
    +\, y^2t^3  ( 1 - 6\rho\eta^2 + 4\rho\eta^3 - 2\rho 
              - 6\rho^2\eta - 18\rho^2\eta^2 + 12\rho^3\eta 
              + 2\rho^3 - \rho^4 - 6\eta + 2\eta^3 )
\nonumber  \\&&
    +\, y^2t^4  (  - 2 - 6\rho\eta + 12\rho\eta^2 - 18\rho^2\eta 
                - 2\rho^2  + 4\rho^3 + 6\eta^2 - \eta^3 )
\nonumber  \\&&
    +\, y^2t^5  ( 12\rho\eta - 2\rho - 6\rho^2 + 6\eta - 3\eta^2 )
    + y^2t^6  ( 2 + 4\rho - 3\eta )
    - y^2t^7 
\nonumber  \\&&
    +\, y^2  (  - 2\rho\eta^3 + 2\rho^3\eta^3 - \rho^4\eta^3 + \eta^3 )
    + y^3t  ( 8\rho\eta + \rho\eta^2 + 2\rho\eta^3 - 12\rho^2\eta 
              + \rho^2\eta^2
\nonumber  \\&&
          \qquad     + 3\rho^2\eta^3 + 8\rho^3\eta 
              - \rho^3\eta^2 
              - 2\rho^4\eta - 2\eta - \eta^2 + 3\eta^3 )
\nonumber  \\&&
    +\, y^3t^2  (  - 1 - \rho\eta + 6\rho\eta^2 - 3\rho\eta^3 
              + 4\rho - \rho^2\eta + 9\rho^2\eta^2 - 6\rho^2 
              + \rho^3\eta  + 4\rho^3
\nonumber  \\&&
              \qquad  - \rho^4 + \eta + 9\eta^2 - 3\eta^3 )
\nonumber  \\&&
    +\, y^3t^3  ( 1 + 6\rho\eta - 11\rho\eta^2 - \rho + 9\rho^2\eta 
              - \rho^2 + \rho^3 + 9\eta - 11\eta^2 + \eta^3 )
\nonumber  \\&&
    +\, y^3t^4  ( 3 - 13\rho\eta + 2\rho + 3\rho^2 - 13\eta + 4\eta^2 )
    + y^3t^5  (  - 5 - 5\rho + 5\eta )
    + 2y^3t^6 
\nonumber  \\&&
    +\, y^3  ( 4\rho\eta^2 + \rho\eta^3 - 6\rho^2\eta^2 + \rho^2\eta^3 
          + 4\rho^3\eta^2 - \rho^3\eta^3 - \rho^4\eta^2 - \eta^2 
          - \eta^3 )
\nonumber  \\&&
    +\, y^4t  (  - 6\rho\eta + 2\rho\eta^2 + 6\rho^2\eta + 
             \rho^2\eta^2 - 2\rho^3\eta + 2\eta - 3\eta^2 )
\nonumber  \\&&
    +\, y^4t^2  ( 1 + 4\rho\eta + \rho\eta^2 - 3\rho 
               + 2\rho^2\eta + 3\rho^2  - \rho^3 - 6\eta + 3\eta^2 )
\nonumber  \\&&
    +\, y^4t^3  (  - 3 + 2\rho\eta + 2\rho + \rho^2 + 6\eta - \eta^2 )
    + y^4t^4  ( 3 + \rho - 2\eta )
\nonumber  \\&&
    -\, y^4t^5 
    + y^4  (  - 3\rho\eta^2 + 3\rho^2\eta^2 - \rho^3\eta^2 + \eta^2 ) \ ,
  \\
f_5 &=& 4yt  ( 3 + 5\rho - 5\eta )
       - 40yt^2 
       + 4y  ( 1 + 5\rho\eta - 6\rho + 5\rho^2 + \eta ) + 10y^2t 
\nonumber  \\&&
       -\, 2y^2  (  1  - 5\rho )
       - 6t  (  1 + 10\rho\eta - 5\rho^2 )
       - 2t^2  (  4 + 30\rho - 15\eta ) + 30t^3 
\nonumber  \\&&
       +\, 30\rho^2\eta - 6\eta + 8\eta^2   \  ,
  \\
f_6 &=& 
         yt  (  - 18\rho\eta - 2\rho\eta^2 
             + 6\rho^2\eta - 15\rho^2\eta^2
           + 10\rho^3\eta  + 2\eta + 17\eta^2 - 4\eta^3 )
\nonumber  \\&&
   +\, yt^2  ( 1 - 16\rho\eta + 15\rho\eta^2 - 9\rho - 30\rho^2\eta 
             + 3\rho^2 + 5\rho^3 + 10\eta - \eta^2 )
\nonumber  \\&&
   +\, yt^3  ( 1 + 30\rho\eta - 10\rho - 15\rho^2 + 10\eta - 5\eta^2 )
   + yt^4  ( 7 + 15\rho - 10\eta )
\nonumber  \\&&
   -\, 5yt^5 
   + y  (  - 9\rho\eta^2 + 4\rho\eta^3 + 3\rho^2\eta^2 
          + 5\rho^3\eta^2 + \eta^2 + 8\eta^3 )
\nonumber  \\&&
   +\, y^2t  (  - 1 + 30\rho\eta - 10\rho\eta^2 + 7\rho + 5\rho^2\eta 
           - 11\rho^2 + 5\rho^3 - 11\eta - 4\eta^2 )
\nonumber  \\&&
   +\, y^2t^2  (  - 2 - 25\rho\eta + 18\rho - 19\eta + 5\eta^2 )
   - 15y^2t^3  (  1 + \rho - \eta )
\nonumber  \\&&
   +\, 10y^2t^4 
   + y^2  ( 7\rho\eta + 12\rho\eta^2 - 11\rho^2\eta + 5\rho^2\eta^2 
             + 5\rho^3\eta - \eta - 9\eta^2 )
\nonumber  \\&&
   +\, y^3t  ( 1 - 6\rho + 5\rho^2 + 8\eta )
   + y^3t^2  ( 8 - 5\eta )
   - 5y^3t^3 
\nonumber  \\&&
    +\, y^3  (  - 6\rho\eta + 5\rho^2\eta + \eta )  \ ,
\end{eqnarray}
\begin{eqnarray}
\widetilde{F}_{\pm}&=& \sqrt{\eta}\; {\cal W}_\pm\, 
\left\{ \left[
{K_b} \left( h_{1,\pm} + h_{2,\pm} {{\cal W}_\pm}^2 \right) 
+  {G_b} h_{3,\pm} \right] 
{\delta (z_{\pm})} 
+ {K_b} h_{4,\pm} {\delta^\prime(z_{\pm})} \right\}\ ,
\end{eqnarray}
where
\begin{eqnarray}
h_{1,\pm} &=& 
        -8yt\eta + 8y\eta^2 + 4y^2t + 12y^2\eta + 2y^3t 
              - 2y^3\eta - 2y^4 -16t\eta - 16\eta^2
\nonumber\\&&
       -\, 4\sigma_{\pm}\;  ( 6yt - 2y\eta + 3y^2t + y^2\eta 
              - y^3 - 8t\eta - 4t^2 - 4\eta^2 )
\nonumber\\&&
       -\, 8\sigma_{\pm}^2\;  ( 3yt - y\eta - 5y^2 + 4t + 4\eta )
       \; +\; 16\sigma_{\pm}^3  ( 3y + t + \eta )  \ ,
\\
h_{2,\pm} &=& 
        2\sigma_{\pm}\; y\, ( 8t\eta^2 + 16t^2\eta + 8t^3 - 4yt\eta^2
             + 4yt^3 - 8y^2t\eta - 6y^2t^2 - 2y^2\eta^2 
\nonumber\\&& 
            -\, y^3t^2 + y^3\eta^2 + y^4t + y^4\eta )
         \; +\; 4\sigma_{\pm}^2\; y\, (  - 4t\eta^2 - 8t^2\eta - 4t^3
\nonumber\\&& 
            - 12yt\eta - 8yt^2 - 4y\eta^2 + 2y^2t\eta 
            - y^2t^2 
           +\, 3y^2\eta^2 + 3y^3t + 3y^3\eta )
\nonumber\\&&
       +\, 8\sigma_{\pm}^3\; y\, (  - 4t\eta - 2t^2 - 2\eta^2 
            + 4yt\eta + yt^2 + 3y\eta^2 + 3y^2t + 3y^2\eta )
\nonumber\\&&
       +\, 16\sigma_{\pm}^4\; y\, ( 2t\eta + t^2 + \eta^2 
             + yt + y\eta )  \  ,
\\
h_{3,\pm} &=& 
       - 16yt\eta - 8yt + 8yt^2 + 8y\eta + 8y\eta^2 - 8y^2t 
              + 8y^2\eta + 24t^2 - 24\eta^2
\nonumber\\&&
       +\, 4\sigma_{\pm}\; (  - 10yt - 14y\eta - 5y^2t + 5y^2\eta 
             - 4y^2 + 5y^3 + 12t - 4t^2 + 12\eta
\nonumber\\&&
         +\, 4\eta^2 )
       \; +\; 16\sigma_{\pm}^2\; ( 5y\eta - 2y + 5y^2 - 9t - 9\eta )
       \;+\; 80\sigma_{\pm}^3\; ( y + t + \eta ) \  ,
   \\
h_{4,\pm} &=& 
        4\sigma_{\pm}\; ( 4yt\eta - 4yt^2 + 6y^2t + 2y^2\eta 
             + y^3t - y^3\eta - y^4 - 8t\eta - 8t^2 )
\nonumber\\&&
       +\, 8\sigma_{\pm}^2\; ( 8yt + 4y\eta + y^2t - 3y^2\eta 
             - 3y^3  + 4t\eta + 4t^2 )
\nonumber\\&&
       -\, 16\sigma_{\pm}^3\; ( yt + 3y\eta + 3y^2 - 2t - 2\eta )
       \;-\;  32\sigma_{\pm}^4\;  ( y + t + \eta ) \  ,
\end{eqnarray}
and
\begin{equation}
{\cal W}_{\pm}={1\over\sqrt{y(t+\eta)(2\sigma_{\pm}+y)-y^2t-(t+\eta)^2}}\ ,
\end{equation}
\begin{equation}
{\cal W}={1\over\sqrt{y(t+\eta)(x_0+y)-y^2t-(t+\eta)^2}} \ ,
\end{equation}
\begin{equation}
\sigma _{\pm}=(t-\eta)/(2\tau _{\pm})\ ,
\qquad z_{\pm}=1+t-\rho-y-2\sigma _{\pm}\ .
\end{equation}
The parameters $K_b,G_b$, representing the kinetic energy 
and the chromomagnetic
energy, are defined according to \cite{MW}.
\subsection{Lepton energy distribution}
As regards the HQET correction terms, we only give the energy distribution
in the form of a diagram evaluated numerically. Beneath we also give the
Born level approximation analytically. The analytic formulae for the polarized 
distribution can be simplified if we split the kinematical range of $y$ into
two parts, separated by the value of the charged lepton energy where the 
virtual $W$ boson can stay at rest. This value is
\begin{equation}
y_W = 1 -\sqrt{\rho}+{\eta\over{1-\sqrt{\rho}}}\quad .
\end{equation}
In the formulae below, the superscripts $A,B$ refer to the appropriate
regimes:
\begin{eqnarray}
y < y_W && \qquad {\rm region}\ A \ ,\\
y > y_W && \qquad {\rm region}\ B \ .
\end{eqnarray}
The energy distribution of polarized $\tau$ lepton reads,
\begin{equation}
{{d\Gamma ^{\pm}}\over{dy}}=12{\Gamma}_0 
\left[{1\over 2}f(y)\pm \Delta f(y)\right]\ .
\end{equation}
The function $f(y)$ represents the unpolarized case,
\begin{equation}
f(y)={1\over 6}\zeta ^2 \sqrt{y^2-4\eta}\, \left\{ \zeta 
\left[y^2-3y(1+\eta)+8\eta\right]
+(3y-6\eta)(2-y)\right\}\quad ,
\end{equation}
with
\begin{equation}
\zeta=1-{\rho\over{1+\eta -y}}\quad .
\end{equation}
The function $\Delta f(y)$ reads,
\begin{equation}
\Delta f(y)= {3\over8}\sqrt{\eta|y-1|}\;\phi _1\Psi
\;+\; {1\over 4}\eta\;\phi_2\ ,
\label{eq:deltaf}
\end{equation}
with  
\begin{eqnarray}
\phi _1&=& -5{\lambda}^3/(y-1)^4
  +3{\lambda}(4{\eta}-{\lambda}-{\lambda}^2)/(y-1)^3 +\, 
(4{\eta}{\lambda}-4{\eta}+{\lambda} \nonumber \\
&& +7{\lambda}^2+{\lambda}^3)/(y-1)^2
  +(-1+4{\eta}{\lambda}-28{\eta}+15{\lambda}-{\lambda}^2
-{\lambda}^3)/(y-1)\nonumber \\
&&  -1+12y{\eta}-11y{\lambda}+7y-y^2+12{\eta}{\lambda}
-24{\eta}+14{\lambda}-11{\lambda}^2 \  ,
\end{eqnarray}
\begin{eqnarray}
\phi _2^A&=& \sqrt{y^2-4\eta}\, \left[\, 15{\lambda}^2{\xi}/(y-1)^3 
 +(10{\eta}{\lambda}{\xi}^2-16{\eta}{\xi}+24{\lambda}{\xi}
-10{\lambda}{\xi}^2 \right. \nonumber \\
&&   -20{\lambda}-6{\lambda}^2{\xi})/(y-1)^2
 +(-4-14{\eta}{\lambda}{\xi}^2-48{\eta}{\xi}+66{\eta}{\xi}^2
-24{\eta}{\xi}^3 \nonumber \\
&&  +8{\eta}^2{\xi}^3-76{\lambda}{\xi}+14{\lambda}{\xi}^2+48{\lambda}
  +3{\lambda}^2{\xi}+25{\xi}-26{\xi}^2+8{\xi}^3)/(y-1) \nonumber \\
&& \left.
  +3-3y+57{\eta}{\xi}-22{\eta}{\xi}^2-12{\lambda}{\xi}-21{\lambda}
  +34{\xi}-18{\xi}^2+8{\xi}^3\, \right] \ ,
\end{eqnarray}
\begin{eqnarray}
\phi _2^B&=&15{\zeta}{\lambda}^2/(y-1)^3
 +  ( 60{\eta}{\zeta}{\lambda} - 16
    {\eta}{\zeta} - 30{\eta}{\zeta}^2{\lambda} 
- 16{\zeta}{\lambda} - 21{\zeta}{\lambda}^2\nonumber\\
&&          + 10{\zeta}^2{\lambda} )/(y-1)^2
 +    (  - 104{\eta}{\zeta}{\lambda}
  - 84{\eta}{\zeta} + 52{\eta}{\zeta}^2{\lambda} 
+ 122{\eta}{\zeta}^2 - 40{\eta}{\zeta}^3\nonumber\\
&& + 160{\eta}^2{\zeta} - 160{\eta}^2{\zeta}^2 + 40
         {\eta}^2{\zeta}^3 - 24{\zeta}{\lambda} 
+ 9{\zeta}{\lambda}^2 + 17{\zeta} - 4
         {\zeta}^2{\lambda}- 22{\zeta}^2  \nonumber\\
&& + 8{\zeta}^3 )/(y-1) +  18 - 29y{\eta} 
+ 27y{\lambda} - 21y + 3y^2 + 46{\eta}{\zeta}{\lambda} 
   - 59{\eta}{\zeta} \nonumber\\
&& - 14{\eta}{\zeta}^2{\lambda} + 78{\eta}{\zeta}^2 
- 16{\eta}{\zeta}^3 - 71{\eta}{\lambda} + 59{\eta}
          - 43{\eta}^2{\zeta} + 26{\eta}^2{\zeta}^2 
- 8{\eta}^2{\zeta}^3\nonumber\\
&& + 46{\eta}^2 - 6{\zeta}{\lambda} - 3{\zeta}{\lambda}^2 
+ 69{\zeta} - 6{\zeta}^2{\lambda} 
 - 52{\zeta}^2 + 16{\zeta}^3 - 42{\lambda} + 24{\lambda}^2 \  ,
\end{eqnarray}
where
\begin{equation}
 \xi = 2 - \zeta\ ,\qquad 
\lambda = \rho + \eta \quad .
\end{equation}
The function $\Psi$ can be written in the form,
\begin{equation}
\Psi =\left\{
\begin{array}{ll}
\arccos \omega _{min} \, - \, \arccos \omega _{max}\,  ,\quad & y<1 \\
 \mbox{arcosh}\, \omega _{max} -\mbox{arcosh}\, \omega _{min} \, ,\quad  & y>1
\end{array}\right.   
\end{equation}
with
\begin{equation}
\omega _{min,max}={{2(y-1)t_{min,max}+y(y_m -y)-2\eta}
\over{y\sqrt{(y_m -y)^2+4\eta\rho}}} \quad .
\end{equation}
Due to terms containing inverse powers of $(y-1)$ the expression 
(\ref{eq:deltaf}) for $\Delta f(y)$ is apparently divergent
at $y=1$. However, expanding $\Delta f(y)$ in powers of $(y-1)$
for $y<1$ and $y>1$ one can check that this function is regular
at $y=1$.
\begin{figure}[htb]
\begin{center}
\leavevmode
\epsfxsize = 270pt
\epsfysize = 270pt
\epsfbox[1 140 578 702]{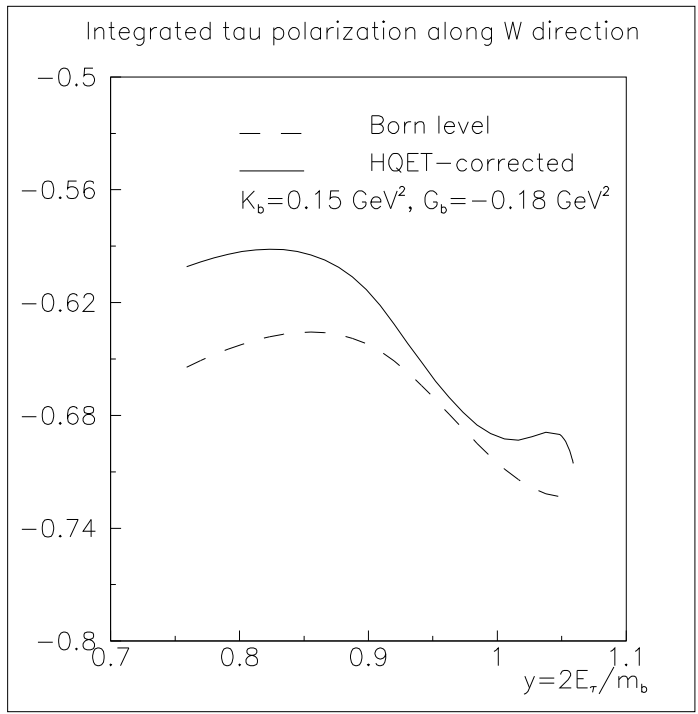}
\end{center}
\caption{Integrated polarization of $\tau$ lepton along the direction 
of virtual W 
in the Born approximation (dashed) and including HQET corrections (solid)
 as functions of the scaled $\tau$ energy $y$. The mass of the $b$ quark
 taken to be $4.75$ GeV, $c$ quark $1.35$ GeV.}
\label{FIG1}  
\end{figure}
\par
The HQET contribution to the decay distributions is known to render them 
unreliable near the endpoint values of the tauon energy. This ambiguity 
reveals itself in the polarization as well. Similar problems appear 
also in calculations of perturbative corrections\cite{JK89,CJ94}.
All these problems are cured if instead of distributions their moments
are considered\cite{CJKK,V95,CJK}. In the case of $\tau$ polarization 
a better defined quantity is 
the integrated polarization
\begin{equation}
P_{int}(\bar y)=
\int _{y_{min}}^{\bar y}\, {dy}\, 
\left(\; {d\Gamma^+\over dy}\; - \; {d\Gamma^-\over dy} \;\right)\; 
\left/ \;
\int _{y_{min}}^{\bar y}\, {dy}\,
\left(\; {d\Gamma^+\over dy}\; + \; {d\Gamma^-\over dy} \;\right) \right.
\label{eq:intpol}
\end{equation}
where both the lowest-order perturbative and the HQET terms are included.
In Fig.\ref{FIG1} the integrated polarization 
is shown as a function of the scaled energy $y$ of the $\tau$ lepton.
The lowest order prediction corresponds to the dashed line and the
solid line is obtained including HQET corrections.
The question arises whether the perturbative QCD
corrections can change this result significantly. As already suggested
in the Introduction, it is plausible that no such thing happens.
\par
On integration over the whole range of the charged lepton energy one 
arrives at 
the total polarization at the tree level corrected for the $O(1/m_b^2)$ 
effects as predicted by HQET. 
For $ m_b = 4.75$ GeV and $ m_c = 1.35$ GeV, we obtain
\begin{equation}
P=-0.7235+4.21{{K_b}\over{m_b^2}}+1.48{{G_b}\over{m_b^2}}\quad .
\end{equation}
Taking $K_b=0.15$ GeV$^2$ and $G_b=-0.18$ GeV$^2$ we obtain $P=-0.706$.
\par
Although we are mostly concerned with the tau lepton polarization here,
the formulae derived in the present work
may well be used in evaluating the polarization of the light
leptons. Interestingly, in the limit of a vanishing mass of the charged 
lepton the polarization falls to zero
apart from the endpoints, c.f. (\ref{eq:deltaf})  and (\ref{LW2}). 
It is due to the chiral $V-A$ structure of the weak charged current that, 
according to Eq.(\ref{Kliniowy}), the decay widths with a definite 
polarization differ by a term proportional to $m_{\tau}s^\mu$. The 
polarization four-vector of the charged lepton can be decomposed as follows: 
\begin{equation}
s^{\mu} = \left(\, s^0,\vec{s}\, \right) =
\left(\,{p\over m}\sqrt{1-(\vec{s}_{\perp})^2},
\vec{s}_{\perp},{E\over m}\sqrt{1-(\vec{s}_{\perp})^2}\,\right)\quad ,
\label{eq:smu}
\end{equation} 
where $\vec{s}_{\perp}$ is understood to mean the part 
of the three-vector $\vec{s}$ perpendicular to the direction
of the charged lepton. The quantities 
$E$ and $p$ denote, respectively, the energy and the three-momentum 
value of the charged lepton. 
This form can easily be seen to meet the definition of 
the polarization four-vector, c.f. Appendix A.
As the lepton mass approaches zero Eq.(\ref{eq:smu})  gives 
\begin{equation}
ms^{\mu}\approx \sqrt{1-(\vec{s}_{\perp})^2}\; {\tau}^{\mu}
\;+\; m\, \left(0,\vec{s}_{\perp},0\right)\quad .
\label{eq:msmu}
\end{equation}
However, if we want to keep the angle subtended by the 
polarization vector and the lepton momentum constant 
the parallel part of the polarization should be proportional
to the perpendicular one, thereby forcing the factor of 
$\sqrt{1-(\vec{s}_{\perp})^2}$ to be of order of $m/E$.
Then the r.h.s of
Eq.(\ref{eq:msmu}) tends to zero for $m\to 0$. 
For the vanishing charged lepton mass the polarization can be 
non-zero only where the virtual $W$ boson is collinear with the charged 
lepton. In general the contribution to polarization 
is appreciable only for $W$ direction within the cone defined by
the condition
\begin{equation}
|\vec{s}_{\perp}|/|\vec s| = {\cal O}(m/E) \quad .
\label{eq:cond}
\end{equation}
In particular this happens if $p$ is much larger than the energy 
of the neutrino. For semitauonic B decays the condition (\ref{eq:cond}) 
is satisfied in the whole phase space and the resulting polarization is
fairly large. 
\section{Summary}
The polarization of the tau lepton along the $W$ boson direction in 
semileptonic $B$ decays has been found at the tree level in perturbative
QCD and the leading order HQET corrections have been included.  The 
quantity is of experimental interest. The fact that it does but slowly 
vary in the regime of low energies of the charged lepton is rather 
favorable in this context \cite{Rozanska}. The QCD one-loop corrections 
are unknown but their irrelevance for the longitudinal polarization 
both in the rest frame of the decaying quark and that of the $W$ boson 
indicates that no great change is to be expected once they are incorporated.
\section{Acknowledgements}
We would like to express our gratefulness to Professor M. R\'o\.za\'nska 
whose suggestion has provided the experimental rationale for this work. 

\appendix
\section{The HQET calculations}
We will presently construct the four-vector $s$ representing a charged 
lepton polarized along the direction of the $W$ boson. 
The defining properties of $s$ are
\begin{equation}\label{s1}
s^2=-1
\end{equation}
and
\begin{equation}\label{s2}
s\cdot\tau=0\ ,
\end{equation}
complemented by the relation $\vec{s} \parallel \vec{W}$. 
Since we are working in the rest frame of the decaying 
meson, the four-vector $s$ can be decomposed as
\begin{equation}\label{sdecomp}
s={\cal A}v+{\cal B}W\ ,
\end{equation}
where $v$ and $W$ stand for the four-velocity of the $B$ 
meson and the four-momentum of the intermediate $W$ boson, 
respectively. While this form automatically satisfies 
$\vec{s}\parallel \vec{W}$, the other two relations 
(\ref{s1}) and (\ref{s2}) have to be imposed, hence yielding 
the expressions for the coefficients appearing in (\ref{sdecomp}). 
With $v=(1,0,0,0)$, one readily identifies:
\begin{equation}
v\cdot\tau=y/2\ ,\qquad v\cdot\nu = x/2\ ,\qquad v^2=-1\ .
\end{equation}
These combined with the other dot products lead to 
the following formulae:
\begin{eqnarray}
{\cal A}_{\pm}&=&{{\mp (t+\eta)}\over{\sqrt{y(t+\eta)(y+x)
-y^2t-(t+\eta)^2}}}\quad ,\\  
{\cal B}_{\pm}&=&{{\pm y}\over{\sqrt{y(t+\eta)(y+x)
-y^2t-(t+\eta)^2}}}\quad .
\end{eqnarray}
The evaluation of the HQET corrections involves differentiation 
over the neutrino energy, once or twice. The denominator in the 
above expressions is easily seen to vanish at the point where 
the $W$ boson is at rest. It is known that the kinematics of 
the process, together with the delta functions and their 
derivatives, finally reduces to integration over the partonic 
phase space. Then there is one such point where the denominator 
vanishes,
\begin{eqnarray}
y&=&1-\sqrt{\rho}+{\eta\over{1-\sqrt{\rho}}}\quad ,\\
t&=& (1 -\sqrt\rho)^2 \quad .
\end{eqnarray}
One might thus raise the question of analyticity of the expressions 
obtained in this way. However, the divergences cancel and moreover 
the resulting distribution is continuous if we ignore the endpoint 
behaviour. That this is so indeed, may be verified by changing the 
variables from $t$ to the square of the three-momentum of the $W$ 
boson. Then the singularity makes its presence only on integration 
over $w_3^2$ rather than affecting the analytical structure of the 
distributions. It turns out that using this variable one obtains an 
analytic expression. This is made clear once one notices that the 
only terms that occur in the course of the calculation are the dot 
products of the four-vector $s$ and the other four-vectors. Writing 
them out explicitly,
\begin{eqnarray}
s_+\cdot v={{\tau _3 \cos \theta}\over{\sqrt{y^2-\tau _3\cos ^2\theta}}}
\quad ,\\
s_+\cdot W={{(y+x)\tau _3\cos\theta 
- 2yw_3}\over{2\sqrt{y^2-\tau _3\cos ^2\theta}}}\quad ,
\end{eqnarray}
with
\begin{equation}
\cos\theta={{w_3^2-\eta-(x-y)/4}\over{w_3\tau _3}}\quad ,
\end{equation}
where the subscript denotes the polarization direction, 
we easily verify that the triple differential distribution 
is analytic in the neutrino energy. 
 Lastly, let us note that another change of variable can be 
useful for evaluating the distribution. Namely, using the 
cosine of the angle subtended by the tau lepton and the 
neutrino\cite{GKS}  eliminates the singular terms from the 
double differential distribution. We have checked numerically 
that the resulting distribution is the same.

\end{document}